\documentclass[conference,twoside]{IEEEtran}
\ifCLASSINFOpdf
\else
\fi

\usepackage{amssymb}
\usepackage{amsthm}
\usepackage[cmex10]{amsmath}
\DeclareMathOperator{\E}{\mathbb{E}}

\DeclareMathOperator{\C}{\mathbb{C}}

\newcommand {\Define} {\stackrel {\Delta} {=}  }
\newcommand{\mya}{\mathrel{\overset{\makebox[0pt]{{\tiny(a)}}}{\approx}}}
\newcommand{\myb}{\mathrel{\overset{\makebox[0pt]{{\tiny(b)}}}{\approx}}}

\newcommand{\myc}{\mathrel{\overset{\makebox[0pt]{{\tiny(c)}}}{\approx}}}
\newcommand{\myd}{\mathrel{\overset{\makebox[0pt]{{\tiny(a)}}}{=}}}

\newcommand{\myg}{\mathrel{\overset{\makebox[0pt]{{\tiny(a)}}}{\ll}}}

\usepackage{bm}
\usepackage{mathtools}
\usepackage{fixltx2e}
\usepackage{epsfig,makeidx,color}
\usepackage{graphicx,dblfloatfix}
\usepackage{amsbsy}
\usepackage{amssymb}
\usepackage{euscript}
\usepackage{lipsum}
\usepackage{cite}
\usepackage{chngcntr}
\usepackage{lineno}
\usepackage[T1]{fontenc}
\usepackage{microtype}
\usepackage{wasysym}
\newtheorem{theorem}{Theorem}

\usepackage[english]{babel}
    \usepackage[font=small,labelsep=space]{caption}
\IEEEoverridecommandlockouts
\begin{document}
%
\title{Low-Complexity CFO Estimation for Multi-User Massive MIMO Systems}
%
%
%
\author{\IEEEauthorblockN{Sudarshan Mukherjee and Saif Khan Mohammed}
\IEEEauthorblockA{ \thanks{The authors are with the Department of Electrical Engineering, Indian Institute of Technology Delhi (IITD), New delhi, India. Saif Khan Mohammed is also associated with Bharti School of Telecommunication Technology and Management (BSTTM), IIT Delhi. Email: saifkmohammed@gmail.com. This work is supported by EMR funding from the Science and Engineering
Research Board (SERB), Department of Science and Technology (DST),
Government of India.}
\thanks{The authors would like to thank Mr.~Indra Bhushan for the initial feasibility study of CFO estimation algorithms in frequency flat massive MIMO channel.}}
}
\maketitle

\begin{abstract}
Low-complexity carrier frequency offset (CFO) estimation and compensation in multi-user massive multiple-input multiple-output (MIMO) systems is a challenging problem. The existing CFO estimation algorithms incur tremendous increase in complexity with increasing number of base station (BS) antennas, $M$ and number of user terminals (UTs) $K$ (i.e. massive MIMO regime). In this paper, we address this problem by proposing a novel low-complexity algorithm for CFO estimation which uses the pilot signal received at the BS during special uplink slots. The total per-channel use complexity of the proposed algorithm increases only linearly with increasing $M$ and is independent of $K$. Analysis reveals that the CFO estimation accuracy can be considerably improved by increasing $M$ and $K$ (i.e. massive MIMO regime). For example, for a fixed $K$ and a fixed training length, the required per-user radiated power during uplink training decreases as $\frac{1}{\sqrt{M}}$ with increasing $M$.
\end{abstract}


%
\IEEEpeerreviewmaketitle

\section{Introduction}
%
%
%
%

Massive multiple-input multiple-output (MIMO)/large scale antenna system (LSAS) has emerged as one of the possible key technologies for the next generation $5$G cellular wireless network because of significant improvement in energy and spectral efficiency over conventional communication technologies \cite{Andrews}. In massive MIMO systems the base station (BS) is equipped with a large number of antennas (in hundreds) to simultaneously serve few tens of autonomous user terminals (UTs) in the same time-frequency resource \cite{Marzetta1}. Increasing the number of BS antennas, $M$, provides array gain, which reduces the energy requirement for transmission \cite{Marzetta1,Ngo1}. These results however assume coherent operation between the UTs and the BS. In practice, the carrier frequency offsets (CFOs) between the carrier frequency of the signal, received at the BS and the BS oscillator can severely impact coherent detection of data, thus degrading the system performance. Substantial amount of work on CFO estimation and optimal pilot design for the same in conventional MIMO systems has been carried out in the past decade \cite{Stoica,Ma,Simon, Ghogho, Poor}. However because of their prohibitively increasing complexity with the number of BS antennas and UTs, these algorithms are not suitable for implementation in massive MIMO systems. In \cite{Larsson2}, for multi-user massive MIMO systems, the authors propose an approximation to the joint ML estimator for the CFOs of all the UTs. The approximation still requires a multi-dimensional grid search and is therefore expected to have high complexity for large number of users. Also, \cite{Larsson2} only considers frequency-flat channel.


\par In this paper, we address the problem of CFO estimation in a frequency-selective multi-user massive MIMO system operating in time divison duplexed (TDD) mode. The contributions of the work presented in this paper are: (i) we propose a simple uplink training scheme which satisfies the optimality criterion for pilot design \cite{Stoica, Ma} (with respect to minimizing the Cramer-Rao Lower Bound (CRLB) on CFO estimation); (ii) we propose a low-complexity multi-user CFO estimation algorithm at the BS and derive analytical expression for its mean square error (MSE). Exhaustive numerical simulations reveal that the proposed CFO estimator performs close to the CRLB, for sufficiently high per-user radiated power; (iii) analysis of the MSE expression reveals that for a fixed $K$ and fixed training length, the required per-user radiated power to achieve a fixed desired MSE decreases as $1/\sqrt{M}$ with increasing $M$, provided that $M$ is sufficiently large (i.e., massive MIMO regime); (iv) for a fixed and sufficiently large $M$ (i.e. massive MIMO regime) and fixed training length, the required per-user radiated power to achieve a fixed desired MSE decreases with increasing $K$ (as long as $K$ is much smaller than the training length); and (v) the per-channel use total complexity of the proposed estimator is $\mathcal{O}(M)$ and is independent of the number of the UTs, $K$. Therefore the required per-user radiated power to achieve a fixed desired MSE can be reduced by increasing $K$ (fixed $M$ and fixed training length), without any increase in the total per-channel use complexity. To the best of our knowledge, for frequency-selective massive MIMO systems, there is no other work which has analytically studied the impact of increasing number of BS antennas and number of UTs on the required per-user radiated power to achieve a fixed desired MSE.

\indent \textbf{{Notations:}} $\C$ denotes the set of complex numbers. $\E$ denotes the expectation operator. $(.)^{H}$ denotes the complex conjugate transpose operation, while $(.)^{\ast}$ denotes the complex conjugate operator. Also, $\bm I_{N}$ denotes the $N\times N$ identity matrix. 



\section{System Model}

We consider a massive MIMO system with $M$ base station (BS) antennas, communicating simultaneously with $K$ autonomous single-antenna user terminals (UTs) in the same time-frequency resource. Massive MIMO systems are expected to operate in Time Division Duplexed (TDD) mode, where each coherence interval is divided into a uplink (UL) slot, followed by a downlink (DL) slot. In the UL, the BS uses pilot signals transmitted by UTs for channel estimation. These channel estimates are used to facilitate coherent detection of user data at the BS receiver. In the DL, the same channel estimates are used to beamform information symbols to the UTs.

\par
\begin{figure}[t]
\includegraphics[width= 3.5 in, height= 1.7 in]{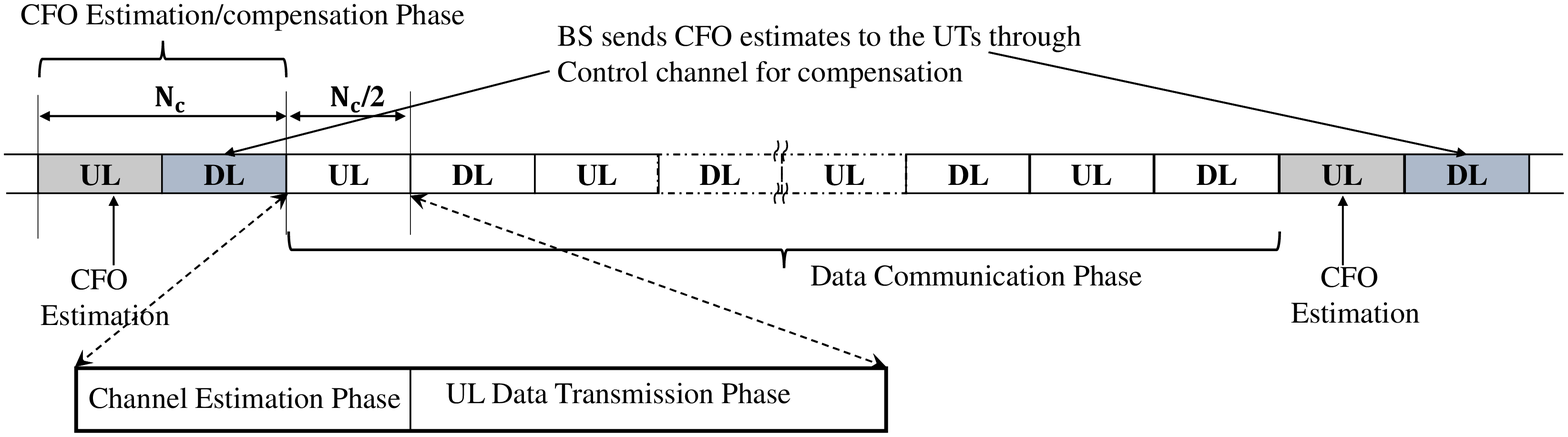}
\caption {The communication strategy, depicting allocation of separate UL/DL TDD slots for CFO estimation and compensation prior to data communication (shaded slots). For UL data communication, the UL slot spans half of the coherence interval, i.e., $\frac{N_c}{2}$ channel uses. The channel estimates acquired in a UL slot are used for downlink beamforming of information to the UTs, in the next DL slot.}
\label{fig:uplink}
\end{figure}

 In a massive MIMO BS serving several UTs simultaneously, acquisition and compensation of carrier frequency offsets from different UTs is important. Therefore we propose the following communication strategy, as illustrated in Fig.~\ref{fig:uplink}. At the beginning  of communication, we propose to first estimate the CFOs from different UTs at the BS, using special pilots transmitted by the UTs in a special UL slot. In the following DL slot, the BS feeds the CFO estimates back to the UTs over a control channel\footnote[1]{We consider this feedback to be error free. Exact mechanism for this feedback and impact of feedback errors is a topic of future study.} and the UTs correct their internal oscillators accordingly (see Fig.~\ref{fig:uplink}). From the next UL slot, normal communication starts, and the UTs transmit pilot signals for channel estimation, followed by uplink data transmission. The special UL/DL slots for CFO estimation might be repeated after several coherence intervals, depending on how fast the CFO changes.

\par The massive MIMO BS under consideration is assumed to operate in a frequency-selective channel environment. Therefore, the complex baseband discrete time channel impulse response has $L>1$ taps. The channel gain coefficient between the $k^{\text{th}}$ UT and the $m^{\text{th}}$ BS antenna at the $l^{\text{th}}$ tap is given by $h_{km}[l] \Define \sigma_{hkl}\hspace{0.1 cm} g_{km}[l]$ ($\sigma_{hkl} > 0$), where $l = 0, 1, \ldots, L-1$, $k = 1, 2, \ldots, K$ and $m = 1, 2, \ldots, M$. Here $g_{km}[l]$ models the fast fading component of the baseband channel gain and are assumed to be independent and identically distributed (i.i.d.) and Rayleigh faded, i.e., circular symmetric complex Gaussian with unit variance, or $g_{km}[l] \sim \mathcal{C}\mathcal{N}(0,1)$. Also, $\{\sigma_{hkl}^2\}, \forall (k,l)$ models the power delay profile (PDP) of the channel. For communication purposes, we assume that the PDP is fixed for the entire duration of communication and is known to the BS. Also the channel realization remains unchanged over $N_c$ channel uses.\footnote[2]{Here $N_c$ is the number of channel uses that span the coherence interval. If $T_c$ is the coherence time and $B_{\text{w}}$ is the communication bandwidth, then $N_c = T_c B_{\text{w}}$.}

\subsection{Signal Model}

Let $a_k[t]$ denote the transmitted signal from the $k^{\text{th}}$ UT at time $t$. The signal received at the $m^{\text{th}}$ BS antenna is therefore given by\footnote[3]{We assume a collocated BS model where all the BS antennas use the same oscillator.}

\vspace{-0.7 cm}

\begin{IEEEeqnarray}{rCl}
\label{eq:rxtsig}
r_m[t] \Define \sum\limits_{k = 1}^{K}\sum\limits_{l = 0}^{L-1}h_{km}[l]a_k[t-l]e^{j\omega_k t} + n_m[t],
\IEEEeqnarraynumspace
\end{IEEEeqnarray}

\noindent where $n_m[t] \sim \mathcal{C}\mathcal{N}(0,\sigma^2)$ is the circular symmetric additive white Gaussian noise (AWGN) with variance $\sigma^2$, $\omega_k \Define 2\pi \Delta f_k T_s$ where $\Delta f_k$ is the frequency offset at the BS for the $k^{\text{th}}$ UT and $T_s \Define 1/B_{\text{w}}$. Here $B_{\text{w}}$ is the total communication bandwidth.

\section{CFO Estimation}

In this section, we first present the Cramer-Rao Lower Bound (CRLB) on the accuracy of the best possible unbiased CFO estimate (see Section III-A). The CRLB also gives us insight into the optimal pilot signal design. Various algorithms have been devised for CFO estimation in single and multi-carrier MIMO systems\cite{Ghogho, Ma, Simon,Poor}. However because of prohibitive increase in their computational complexity with increasing number of BS antennas and UTs, these algorithms are not suitable for massive MIMO systems. We propose a simple orthogonal pilot sequence, which satisfies the optimality criterion for pilot design (see Section III-B). Using these pilots we develop a \textit{low-complexity} algorithm to estimate the CFO of each user separately (Section III-C).

\subsection{CRLB for CFO Estimation}

Let $(a_k[0], a_k[1], \cdots, a_k[N-1])$ be the sequence of $N$ pilot symbols transmitted by the $k^{\text{th}}$ UT in the UL slot for CFO estimation (see Fig.~\ref{fig:uplink}).\footnote[4]{A copy of the last $L-1$ pilot symbols, i.e., $\{a_k[N-L+1], \cdots, a_k[N-1]\}$ is transmitted before $\{a_k[0], \cdots, a_k[N-1]\}$.} The signal received at the $m^{\text{th}}$ BS antenna over the $N$ channel uses is denoted by $\bm r_m \Define (r_m[0], r_m[1], \cdots, r_m[N-1])^T$. Using \eqref{eq:rxtsig}, with $t = 0, 1, \ldots, N-1$ (here $t$ represents the $t^{\text{th}}$ channel use), $\bm r_m$ is given by \cite{Ma}

\vspace{-0.7 cm}

\begin{IEEEeqnarray}{rCl}
\label{eq:rxsigv}
\bm r_m & = & \sum\limits_{k = 1}^{K}\bm \Gamma(\omega_k) \bm A_k \bm h_{km} + \bm n_{m}.
\IEEEeqnarraynumspace
\end{IEEEeqnarray}

\vspace{-0.2 cm}

\indent Here, $\bm \Gamma(\omega_k) \Define diag(1, e^{j \omega_k}, e^{j2\omega_k}, \cdots, e^{j (N-1)\omega_k})$, $\bm n_m \Define (n_m[0], n_m[1], \cdots, n_m[N-1])^T$, and $\bm h_{km} \Define \big(h_{km}[0], h_{km}[1], \cdots, h_{km}[L-1]\big)^T$. Also $\bm A_k \in \C ^{N \times L}$ is a circulant matrix whose $(t,q)^{\text{th}}$ entry is given by

\vspace{-0.3 cm}

\begin{IEEEeqnarray}{rCl}
\label{eq:sigk}
A_k(t, q) = 
\left\{ \begin{array}{ll}
a_k[t-q] & t \geq q\\
a_k[t-q+N] & t < q
\end{array}\right.
\IEEEeqnarraynumspace
\end{IEEEeqnarray}

\noindent where $t = 1, 2, \ldots, N$, and $q = 1, 2, \ldots, L$. The overall received signal vector, $\bm r \Define (\bm r_1^T, \bm r_2^T, \cdots, \bm r_M^T)^T$, is thus given by \cite{Ma}

\vspace{-0.7 cm}

\begin{IEEEeqnarray}{rCl}
\label{eq:rxsig}
\bm r = \bm Q(\bm \omega)\bm h + \bm n,
\IEEEeqnarraynumspace
\end{IEEEeqnarray}

\noindent where $\bm h \Define (\bm h_1^T, \bm h_2^T, \cdots, \bm h_M^T)^T$ and $\bm h_m \Define (\bm h_{1m}^T, \bm h_{2m}^T, \cdots, \bm h_{Km}^T)^T$. Further, $\bm n \Define (\bm n_1^T, \bm n_2^T, \cdots, \bm n_M^T)^T$, and $\bm \omega \Define (\omega_1, \omega_2, \cdots, \omega_K)^T$. Also, $\bm Q(\bm \omega)  \Define  \bm I_M \otimes [\bm \Gamma(\omega_1)\bm A_1, \bm \Gamma(\omega_2)\bm A_2, \cdots, \bm \Gamma(\omega_K)\bm A_K]$. Here `$\otimes$' denotes the Kronecker product for matrices.

\par Based on the above data model, the CRLB for CFO estimation is given by \cite{Stoica, Ma}:

\vspace{-0.3 cm}

\begin{IEEEeqnarray}{lCl}
\label{eq:crlb}
\text{CRLB}(\bm \omega) & \Define &  \dfrac{\sigma^2}{2}\bigg(\Re \bigg\{\bm V^H\bm P_Q \bm V\bigg\}\bigg)^{-1},
\IEEEeqnarraynumspace
\end{IEEEeqnarray}

\vspace{-0.2 cm}

\noindent where $\bm P_Q \Define  \bm I_{MN} - \bm Q(\bm \omega)( \bm Q^H(\bm \omega) \bm Q(\bm \omega))^{-1}\bm Q^H(\bm \omega)$. In \eqref{eq:crlb}, the submatrix of $\bm V$ defined by rows $[(m-1)N+1, \ldots, mN]$ in the $k^{\text{th}}$ column is given by $\bm V((m-1)N+1:mN,k) \Define \bm F\bm \Gamma(\omega_k)\bm A_k\bm h_{km} \in \C^{N\times 1}$, where $\bm F = diag(0, 1, \cdots, N-1)$, $ m = 1, 2, \ldots, M$, and $k = 1, 2, \ldots, K$.

\begin{figure}[t]
\includegraphics[width= 3.5 in, height= 1.2 in]{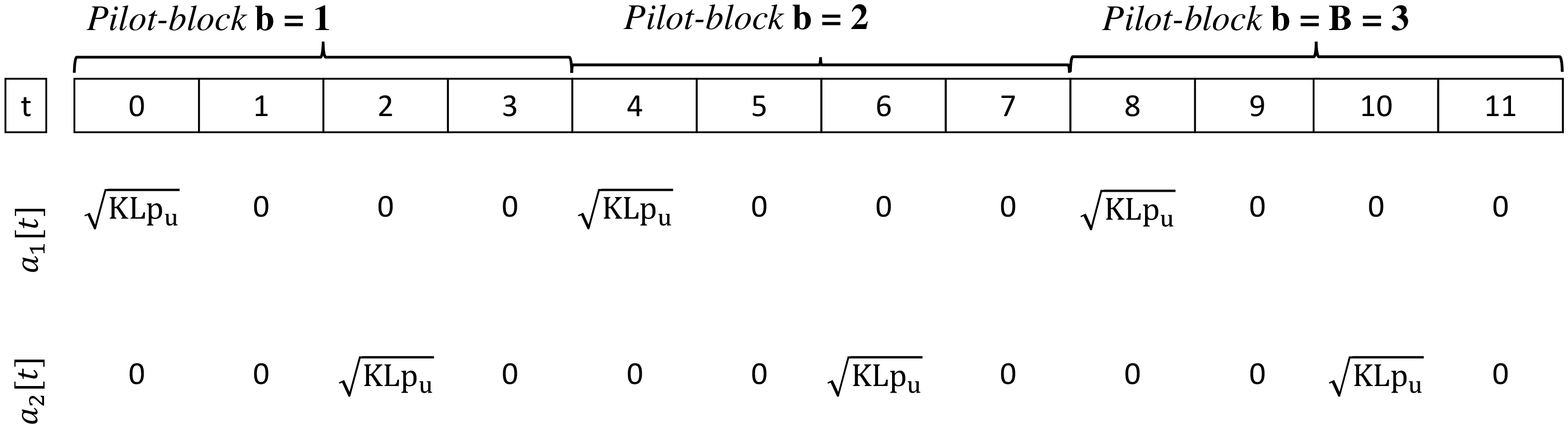}
\caption {Illustration of the proposed transmitted pilot/training sequence for CFO estimation with $K = 2$, $L = 2$, $N = 12$ (i.e., $B = N/KL = 3$).}
\label{fig:pilot}
\end{figure}

\subsection{Proposed Training Sequence for CFO Estimation}

From \cite{Stoica, Ma} for large $N$, the conditions for optimality of the training sequence (i.e. in terms of minimizing the CRLB) is given by

\vspace{-0.3 cm}

\begin{IEEEeqnarray}{lCl}
\label{eq:optpilot}
\left. \begin{array}{lll}
\bm A_i^H\bm A_i & \propto & \bm I_{L}\\
\bm A_i^H\bm A_j & = & \bm 0_{L\times L},
\end{array} \right\} \forall i \neq j, (i,j) \in \{1, 2, \ldots, K\}.
\IEEEeqnarraynumspace
\end{IEEEeqnarray}

\indent In this paper, we propose pilot signals which consist of \textit{pilot-blocks}. Each \textit{pilot-block} spans $KL$ channel uses and there are therefore $B \Define \frac{N}{KL}$ \textit{pilot-blocks}. In each \textit{pilot-block}, each UT transmits only a single impulse, which is preceded and followed by zeros. In the $b^{\text{th}}$ \textit{pilot-block}, the first UT transmits an impulse at $t = (b-1)KL$, the second UT at $t = (b-1)KL +L$, and the $K^{\text{th}}$ UT at $t = (b-1)KL + (K-1)L$. Since the impulses from different UTs are separated by $L$ channel uses (i.e. the maximum delay spread of any user's impulse response), the signal received at the BS antennas in the time interval $(b-1)KL + (k-1)L \leq t \leq (b-1)KL +(kL-1)$ will simply be the impulse response of the channel for the $k^{\text{th}}$ UT. These $KL$ channel uses ($L$ channel use for each of the $K$ users) constitute a \textit{pilot-block} (see Fig.~\ref{fig:pilot}). Since the channel realization remains static over an entire coherence interval, the CFO of a user can be estimated by correlating the impulse responses received from that user in consecutive \textit{pilot-blocks}. The overall transmitted pilot for the $k^{\text{th}}$ user is given by

\vspace{-0.3 cm}

\begin{IEEEeqnarray}{rCl}
\label{eq:kpilotmat}
\bm a_k[t] = \left\{
\begin{array}{ll}
\sqrt{KLp_{\text{u}}}, & t\, \text{mod}\, KL = (k - 1)L\\
0, & \text{elsewhere},
\end{array}\right.
\IEEEeqnarraynumspace
\end{IEEEeqnarray}

\noindent where $p_{\text{u}}$ is the average power of the pilot signal transmitted by each UT.\footnote[5]{Use of the scaling factor $\sqrt{KL}$ guarantees that the average power transmitted from each user is $p_{\text{u}}$, irrespective of the number of users and the number of channel taps.}  An example of the proposed pilot sequence is illustrated in Fig.~\ref{fig:pilot}.

\par It can be shown that for the proposed training sequence in \eqref{eq:kpilotmat}, $\bm A_i^H \bm A_j = \bm 0$ and $\bm A_i^H\bm A_i = KL \bm I_L$, when $i \neq j$, $i = 1, 2, \ldots, K$ and $j = 1, 2, \ldots, K$, i.e., the proposed training sequence satisfies the optimality conditions (see \eqref{eq:optpilot}).

\begin{figure*}[!t]
\normalsize
\vspace*{-20pt}
\begin{IEEEeqnarray}{lCl}
\nonumber r_m^{\ast}[\tau(b,k,l)]r_m[\tau(b+1,k,l)]   =   KLp_{\text{u}}|h_{km}[l]|^2e^{j\omega_k KL} +\\
 \hspace{3.3 cm} \underbrace{\stackrel{\bigg[\underbrace{\stackrel{\sqrt{KLp_{\text{u}}}h_{km}^{\ast}[l]e^{-j\omega_k \tau(b,k,l)}n_m[\tau(b+1,k,l)]}{} }_{\Define \displaystyle T_1} + \underbrace{\stackrel{\sqrt{KLp_{\text{u}}}h_{km}[l]e^{j\omega_k \tau(b+1,k,l)}n_m^{\ast}[\tau(b,k,l)]}{}}_{\Define \displaystyle T_2}\bigg] + \underbrace{\stackrel{n_m[\tau(b+1,k,l)]n_m^{\ast}[\tau(b,k,l)]}{}}_{\Define \displaystyle T_3}}{}}_{\text{Sum of Noise Terms, } \displaystyle c_k(b,m,l) \Define T_1 + T_2 + T_3.}.
 \IEEEeqnarraynumspace
\label{eq:eqnintmed1}
\end{IEEEeqnarray}

\hrulefill
\end{figure*}

\subsection{Proposed CFO estimate for the $k^{\text{th}}$ UT}

For the proposed pilot signal given in \eqref{eq:kpilotmat}, the received signal at the $m^{\text{th}}$ BS antenna at time instance $\tau(b,k,l) \Define (b-1)KL + (k-1)L +l$ is given by

\vspace{-0.45 cm}

\begin{IEEEeqnarray}{rCl}
\label{eq:rxpilotcfo1}
\nonumber r_m[\tau(b,k,l)] = \sqrt{KL p_{\text{u}}} \hspace{0.1 cm} h_{km}[l] \hspace{0.1 cm} e^{j\omega_k\tau(b,k,l)} + n_m[\tau(b,k,l)],
\IEEEeqnarraynumspace
\end{IEEEeqnarray}

\noindent where $l = 0, 1, \ldots, L-1$, $k = 1, 2, \ldots, K$ and $ b = 1, 2, \ldots, B$. The correlation of the received pilots in consecutive blocks, i.e., at time instances $\tau(b,k,l)$ and $\tau(b+1,k,l)$ is given by \eqref{eq:eqnintmed1} (see top of the next page). Note that the argument of the first term in \eqref{eq:eqnintmed1} is $\omega_k KL$, i.e., it depends on the CFO of the $k^{\text{th}}$ user, $\omega_k$. Since the argument of this term is independent of the BS antenna index, $m$, the channel tap $l$ and the \textit{pilot-block} index $b$, we can average the noise terms in \eqref{eq:eqnintmed1} (denoted as $c_k(b,m,l)$) over $b$, $m$ and $l$. This averaging is given by

\vspace{-0.3 cm}

\begin{IEEEeqnarray}{rCl}
\label{eq:rhok}
\nonumber \rho_k & \Define & \sum\limits_{b = 1}^{B-1} \sum\limits_{m = 1}^{M} \sum\limits_{l = 0}^{L-1} \dfrac{r_m^{\ast}[\tau(b,k,l)]r_m[\tau(b+1,k,l)]}{MKL(B-1)p_{\text{u}}\sum\limits_{l = 0}^{L -1}\sigma_{hkl}^2}\\
\label{eq:rhokredef}
 & = & G_ke^{j\omega_k KL} + \nu_k.
\IEEEeqnarraynumspace
\end{IEEEeqnarray}

\vspace{-0.2 cm}

\indent Here $\nu_k$ and $G_k$ are given by

\vspace{-0.5 cm}

\begin{IEEEeqnarray}{rCl} 
\label{eq:cfonoise}
\nu_k & \Define & \dfrac{\sum\limits_{b=1}^{B-1} \sum\limits_{m=1}^{M} \sum\limits_{l=0}^{L-1} c_k(b,m,l)}{MKL(B-1)p_{\text{u}}\sum\limits_{l=0}^{L-1}\sigma_{hkl}^2},\\
%
%
%
\label{eq:gk}
G_k & \Define & \left(\dfrac{\sum\limits_{m = 1}^{M} \sum\limits_{l = 0}^{L -1}  |h_{km}[l]|^2}{M \sum\limits_{l = 0}^{L-1}\sigma_{hkl}^2}\right).
\IEEEeqnarraynumspace
\end{IEEEeqnarray}

\par From \eqref{eq:rhokredef} we note that the argument of the first term depends on the CFO of the $k^{\text{th}}$ UT and its magnitude, $G_k$, is independent of the CFO. Therefore, we propose the following estimate of the CFO for the $k^{\text{th}}$ user

\vspace{-0.6 cm}

\begin{IEEEeqnarray}{c}
\label{eq:cfoest}
\hat{\omega}_k \Define \dfrac{\arg{(\rho_k)}}{KL},
\IEEEeqnarraynumspace
\end{IEEEeqnarray}

\vspace{-0.2 cm}

\noindent where $\arg(c)$ denotes the `principal argument' of the complex number $c$.

\textit{Remark 1:\label{ppm}} Note that the proposed CFO estimate in \eqref{eq:cfoest} is well-defined if and only if $|\omega_k KL| < \pi$. For most practical systems, we believe that this condition will hold true.\footnote[6]{For instance, with a carrier frequency of $f_c = 2$ GHz, and a oscillator accuracy of $0.1$ PPM (parts per million) (commonly used in cellular BSs \cite{Weiss}) the maximum carrier frequency offset is $\Delta f = f_c \times 10^{-7} = 200$ Hz. For a system having communication bandwidth $B_{\text{w}} = 1$ MHz (i.e., $T_s = 10^{-6}$ s) this corresponds to a CFO of $\omega_k = 2\pi \Delta f T_s = 2\pi \times 200 \times 10^{-6} = 4\pi \times 10^{-4}$ radian, which is 2500 times less than $\pi$. Therefore for massive MIMO systems, even with $K = 10$ and $L = 5$, $|\omega_k KL| = \frac{\pi}{50} \ll \pi$.} \hfill \qed

\textit{Remark 2:\label{maxK}} With a carrier frequency $f_c$, let the maximum CFO for any user be $\Delta f = \kappa f_c$ (note that for mobile terminals, $\kappa$ might depend on the velocity of the terminal.). Since $|\omega_k KL|$ must be less than $\pi$ and $\omega_k = 2\pi \Delta f T_s$, we must have

\vspace{-0.5 cm}

\begin{IEEEeqnarray}{lCl}
\label{eq:maxk}
\pi & > & |2\pi \Delta f KL T_s| \myd  2\pi \kappa Kf_c T_\text{d},
\IEEEeqnarraynumspace
\end{IEEEeqnarray}

\vspace{-0.2 cm}

\noindent where $(a)$ follows from the fact that $T_\text{d} \Define LT_s$ is the delay spread of the channel. Therefore from \eqref{eq:maxk}, it follows that, for the proposed CFO estimator in \eqref{eq:cfoest} to work, the maximum number of allowed users must be less than $\dfrac{1}{2\kappa f_c T_{\text{d}}}$. Note that this maximum limit is usually quite large. For example with $f_c = 2$ GHz, $T_{\text{d}} = 5 \mu$s and $\kappa = 0.1$ PPM, we have $K < 500$. \hfill \qed

\textit{Remark 3:\label{complexity}} From \eqref{eq:rhokredef} and \eqref{eq:cfoest} it is clear that the total number of operation required to compute all the $K$ CFO estimates is $(M(B-1)L+1)K$. Since $N= BKL$, the average number of operations per-channel use is $(M(B-1)L+1)K/N \approx M$ ($N \gg K$). Note that the total per-channel use complexity is independent of $K$ and increases only linearly with $M$. \hfill \qed

\textit{Remark 4:\label{gkavg}} From \eqref{eq:gk} and the strong law of large numbers, it follows that for i.i.d. $\{h_{km}[l]\}$, $G_k \to 1$ as $M \to \infty$, with probability $1$. \hfill \qed

\textbf{Lemma 1\label{noisevar}}. The mean and variance of $\nu_k^I \Define \Re(\nu_k)$ and $\nu_k^Q \Define \Im(\nu_k)$ are given by $\E[\nu_k^I] = \E[\nu_k^Q] = 0$ and

\vspace*{-0.3 cm}

\begin{IEEEeqnarray}{lCl}
\label{eq:vkvarI}
\nonumber \E[(\nu_k^I)^2]   =   \frac{\frac{G_k}{\gamma_k}\Big(1 + \frac{B-2}{B-1}\cos(2\omega_k KL)\Big)}{MKL(B-1)} +\frac{\frac{1}{\displaystyle 2K\gamma_k^2}}{MKL(B-1)}, \\
\IEEEeqnarraynumspace
\end{IEEEeqnarray}
\vspace{-0.8 cm}
\begin{IEEEeqnarray}{lCl}
\label{eq:vkvarQ}
\nonumber \E[(\nu_k^Q)^2]  =  \frac{\frac{G_k}{\gamma_k}\Big(1 - \frac{B-2}{B-1}\cos(2\omega_k KL)\Big)}{MKL(B-1)} +\frac{\frac{1}{\displaystyle 2K\gamma_k^2}}{MKL(B-1)}.\\
\IEEEeqnarraynumspace
\end{IEEEeqnarray}

\vspace{-0.3 cm}
\indent Here $\gamma_k \Define \dfrac{p_{\text{u}}}{\sigma^2}\sum\limits_{l = 0}^{L-1}\sigma_{hkl}^2$. 

\vspace{0.1 cm}
\begin{IEEEproof}
From \eqref{eq:cfonoise} we have

\vspace{-0.4 cm}

\begin{IEEEeqnarray}{rCl}
\label{eq:lemmaeqn2}
\vspace{-0.15 cm}
\nonumber \nu_k^I & \Define & \Re(\nu_k) = \dfrac{1}{MKL(B-1)p_{\text{u}}\sum\limits_{l=0}^{L-1}\sigma_{hkl}^2} \Re\{X_k\},\\
\vspace{-0.5 cm}
\nu_k^Q & \Define & \Im(\nu_k) = \dfrac{1}{MKL(B-1)p_{\text{u}}\sum\limits_{l=0}^{L-1}\sigma_{hkl}^2} \Im\{X_k\},
\IEEEeqnarraynumspace
\end{IEEEeqnarray}

 \noindent where $X_k \Define \sum\limits_{b=1}^{B-1}\sum\limits_{m=1}^{M}\sum\limits_{l=0}^{L-1} c_k(b,m,l)$ and $c_k(b,m,l)$ is defined in \eqref{eq:eqnintmed1}. Since $n_m[t]$ are all complex circular symmetric, i.i.d. Gaussian, we have $\E[T_i] = 0, i = 1, 2, 3$ ($T_i$ are defined in \eqref{eq:eqnintmed1}) and therefore $\E[c_k(b,m,l)] = 0, \forall$ $b = 1, 2, \ldots, B-1$, $l = 0, 1, \ldots, L-1$ and $m = 1, 2, \ldots, M$. 
 \par Clearly, $\E[X_k] = \sum\limits_{b=1}^{B-1}\sum\limits_{m=1}^{M}\sum\limits_{l=0}^{L-1} \E[c_k(b,m,l)] = 0$. Therefore from \eqref{eq:lemmaeqn2}, we get $\E[\nu_k^I] = \E[\nu_k^Q] = \E[\nu_k] = 0$. 
 \par The real components of $T_i$, $i = 1, 2, 3$ are given by
 
\vspace{-0.5 cm}

\begin{IEEEeqnarray}{lCl}
\label{eq:lemmaneqn1}
\nonumber \Re(T_1)  =  \hspace{0.1 cm} \sqrt{KLp_{\text{u}}}|h_{km}[l]| \hspace{0.1 cm} |n_m[\tau(b+1,k,l)]|\\
\nonumber \,\,\,\,\,\,\,\,\,\, \cos \Big(\angle n_m[\tau(b+1,k,l)] - \angle h_{km}[l] - \omega_k \tau(b,k,l)\Big),\\
\nonumber \Re(T_2)  =  \hspace{0.1 cm} \sqrt{KLp_{\text{u}}}|h_{km}[l]|\hspace{0.1 cm} |n_m[\tau(b,k,l)]|\\
\nonumber \,\,\,\, \cos \Big(-\angle n_m[\tau(b,k,l)] + \angle h_{km}[l] + \omega_k \tau(b+1,k,l)\Big),\hspace{-1 cm}\\
\nonumber \text{and, }\,\, \Re(T_3)  =  |n_m[\tau(b,k,l)]|\hspace{0.1 cm} |n_m[\tau(b+1,k,l)]|\\
\,\,\,\,\,\,\,\,\,\,\,\,\, \cos \Big(\angle n_m[\tau(b+1,k,l)] - \angle n_m[\tau(b,k,l)]\Big),
\IEEEeqnarraynumspace
\end{IEEEeqnarray}

\vspace{-0.15 cm}

\noindent where $\angle c$ represents the `principal argument' of $c \in \C$. Using expressions in \eqref{eq:lemmaneqn1}, the variance of $\Re(X_k)$ is given by \eqref{eq:lemmaneqn2} (see top of the next page). Using the expression for $\E[(\Re(X_k))^2]$ from \eqref{eq:lemmaneqn2} and the expression for $\nu_k^I$ from \eqref{eq:lemmaeqn2}, the variance of $\nu_k^I$ is given by \eqref{eq:vkvarI}. Similarly, the variance of $\nu_k^Q$ can be shown to be given by \eqref{eq:vkvarQ}. \hfill \IEEEQEDhere
\end{IEEEproof}

\begin{figure*}[!t]
\normalsize
\vspace*{-20pt}
\begin{IEEEeqnarray}{lCl}
\label{eq:lemmaneqn2}
\nonumber \stackrel{\E[(\Re(X_k))^2]  =  \E\bigg[\bigg(\sum\limits_{b=1}^{B-1}\sum\limits_{m=1}^{M}\sum\limits_{l=0}^{L-1} \Re(T_1 + T_2 + T_3)\bigg)^2\bigg] = \sum\limits_{b=1}^{B-1}\sum\limits_{m=1}^{M}\sum\limits_{l=0}^{L-1}\Biggl[KL p_{\text{u}} |h_{km}[l]|^2 \sigma^2 \bigg\{\E\Big[\cos^2\Big(\omega_k \tau(b+1,k,l)+\angle h_{km}[l] - \angle n_m[\tau(b,k,l)]\Big)\Big]}{}\\
\nonumber \stackrel{ + \E\Big[\cos^2\Big(\angle n_m[\tau(b+1,k,l)] - \omega_k \tau(b,k,l) -\angle h_{km}[l]\Big)\Big]\bigg\} + \sigma^4 \E\Big[\cos^2\Big(\angle n_m[\tau(b+1,k,l)]-\angle n_m[\tau(b,k,l)]\Big)\Big]\Biggl]}{}\\
\nonumber \stackrel{ + 2\sum\limits_{b=2}^{B-1}\sum\limits_{m=1}^{M}\sum\limits_{l=0}^{L-1}\bigg\{KL p_{\text{u}} |h_{km}[l]|^2 \sigma^2 \E\Big[\cos \Big(\angle n_m[\tau(b+1,k,l)] - \angle h_{km}[l] - \omega_k \tau(b+2,k,l)\Big)\cos\Big(\angle n_m[\tau(b+1,k,l)] - \angle h_{km}[l] - \omega_k \tau(b,k,l)\Big)\Big]\bigg\}}{}\\
\stackrel{= \sum\limits_{b=1}^{B-1}\sum\limits_{m=1}^{M}\sum\limits_{l=0}^{L-1}\bigg\{KL p_{\text{u}} |h_{km}[l]|^2 \sigma^2 + \sigma^4/2\bigg\}+ \sum\limits_{b=2}^{B-1}\sum\limits_{m=1}^{M}\sum\limits_{l=0}^{L-1}\bigg\{KL p_{\text{u}} |h_{km}[l]|^2 \sigma^2\cos(2\omega_k KL)\bigg\}}{}.
\IEEEeqnarraynumspace
\end{IEEEeqnarray}
\hrulefill
\end{figure*}

\vspace{0.1 cm}

\textit{Remark 5:\label{noise}} From \eqref{eq:vkvarI} and \eqref{eq:vkvarQ} it is clear that $\E[(\nu_k^I)^2]$ and $\E[(\nu_k^Q)^2]$ decrease with increasing number of BS antennas, $M$. This reduction in the variance of $\nu_k^I$ and $\nu_k^Q$ with increasing $M$ is due to the averaging of the noise terms (defined in \eqref{eq:eqnintmed1}) across all $M$ BS antennas (see \eqref{eq:cfonoise}). Further as $M \to \infty$, since $\nu_k^I$ and $\nu_k^Q$ are average of a large number of noise terms (see \eqref{eq:cfonoise}), it is expected that both $\nu_k^I$ and $\nu_k^Q$ would be asymptotically Gaussian distributed. \hfill \qed

\begin{theorem}
\emph{(Approximation of the CFO Estimate)}
\label{cfoapprox}
If $|\omega_k KL| \ll \pi$ and $\gamma_k \gg \gamma_k^{0}$, then the proposed CFO estimate in \eqref{eq:cfoest} can be approximated by

\vspace{-0.65 cm}

\begin{IEEEeqnarray}{rCl}
\label{eq:approxest1}
\hat{\omega}_k  \approx  \omega_k + \frac{\nu_k^Q}{G_k KL},
\IEEEeqnarraynumspace
\end{IEEEeqnarray}

\vspace{-0.1 cm}

\noindent where $\gamma_k^{0} \Define \dfrac{\frac{B-1}{2B-3}}{KG_k\bigg[\sqrt{1 + 2ML\frac{(B-1)^3}{(2B-3)^2}} - 1\bigg]}$.
\end{theorem}

\begin{IEEEproof}
See Appendix. \hfill  \IEEEQEDhere
\end{IEEEproof}
 
\vspace{0.1 cm}

\begin{figure}[t]
\hspace{-0.3 in}
\includegraphics[width= 3.7 in, height= 2.5 in]{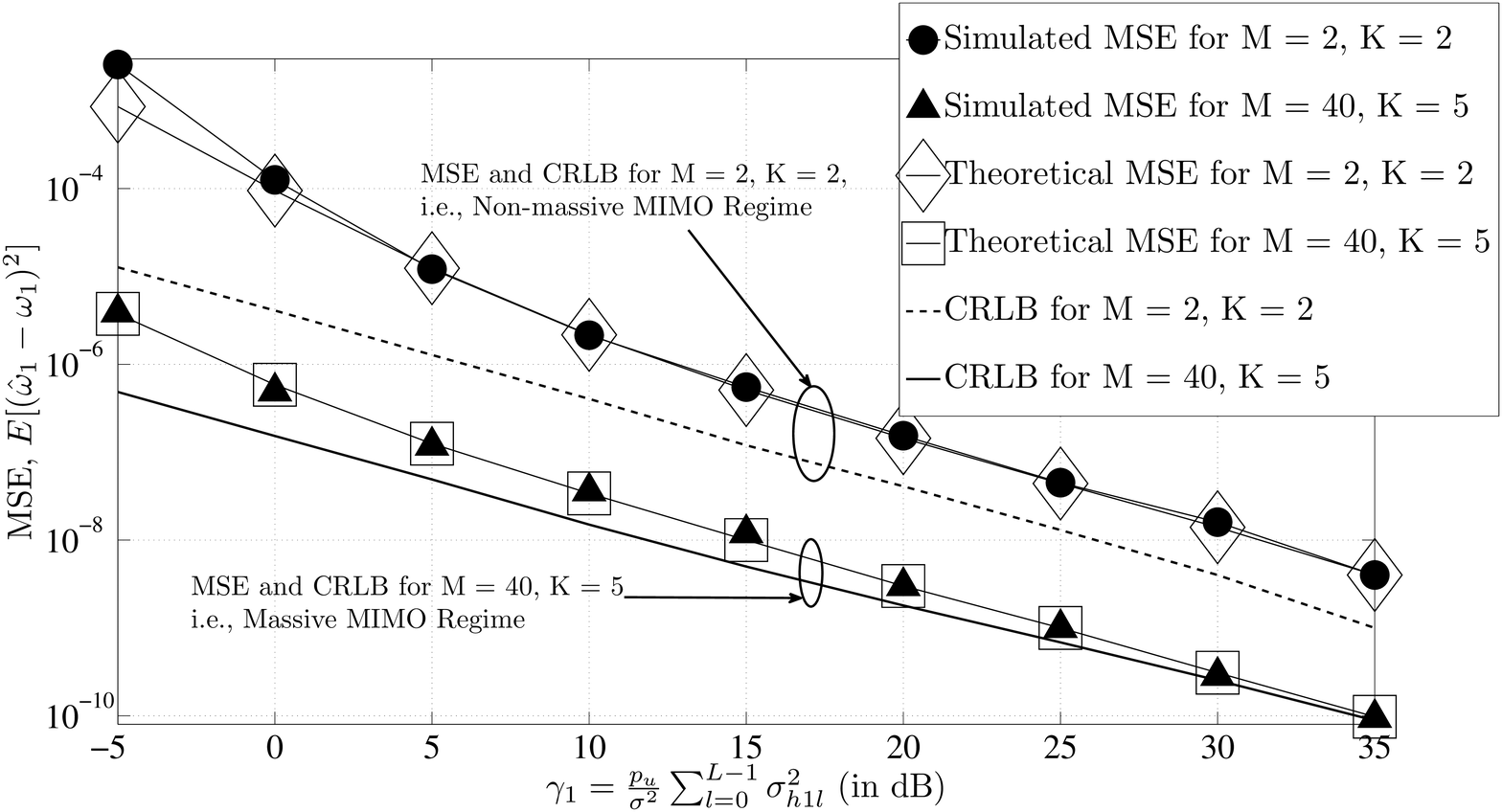}
\caption {Plot of Simulated and Theoretical MSE with varying SNR, compared to the CRLB, depicting the improvement in the MSE in massive MIMO regime ($M = 40, K = 5$), with respect to the non-massive MIMO regime ($M = 2, K = 2$). Fixed $N = 100$ and $L = 2$.}
\vspace{-1 cm}
\label{fig:crlbmse}
\end{figure}

\textbf{Corollary to Theorem 1}: If $|\omega_k KL| \ll \pi$ and $\gamma_k \gg \gamma_k^{0}$, then the mean square error (MSE) of the proposed CFO estimate for the $k^{\text{th}}$ UT is given by

\vspace{-0.5 cm}

\begin{IEEEeqnarray}{rCl}
\label{eq:msewk}
\E[(\hat{\omega}_k - \omega_k)^2] \approx \dfrac{\dfrac{1}{\gamma_k}\bigg(\dfrac{G_k}{B-1} + \dfrac{1}{2K\gamma_k}\bigg)}{M(N-KL)(KL)^2G_k^2}.
\IEEEeqnarraynumspace
\end{IEEEeqnarray}

\begin{IEEEproof}
For $|\omega_k KL| \ll \pi$ and $\gamma_k \gg \gamma_k^{0}$, from \eqref{eq:approxest1}, the expression for CFO estimation error for the $k^{\text{th}}$ UT is given by

\vspace{-0.5 cm}

\begin{IEEEeqnarray}{rCl}
\label{eq:errwk}
\Delta \omega_k \Define \hat{\omega}_k - \omega_k \approx \dfrac{\nu_k^Q}{G_k KL}.
\end{IEEEeqnarray}

\indent Clearly, $\E[\Delta \omega_k] \approx 0$. The variance of $\Delta \omega_k$, i.e., the mean square error (MSE) is then given by

\vspace{-0.5 cm}

\begin{IEEEeqnarray}{rCl}
\label{eq:varerr}
\E[(\Delta \omega_k)^2] & \approx & \dfrac{\E[(\nu_k^Q)^2]}{G_k^2 (KL)^2}.
\IEEEeqnarraynumspace
\end{IEEEeqnarray}

\par Further since $|\omega_k KL| \ll \pi$, we have $\cos(2\omega_k KL) \approx 1$. Using \eqref{eq:vkvarQ} (with the approximation $\cos(2\omega_k KL) \approx 1$) in \eqref{eq:varerr}, we get \eqref{eq:msewk}. \hfill \IEEEQEDhere
\end{IEEEproof}

\textit{Remark 6:\label{mse}} From \eqref{eq:msewk}, it is clear that the MSE diminishes with increasing received SNR, $\gamma_k$. We also illustrate this through Fig.~\ref{fig:crlbmse}, where we plot the variation in the MSE of the CFO for the first user, i.e., $\E[(\hat{\omega}_1 - \omega_1)^2]$ averaged over the channel statistics, as a function of increasing SNR, $\gamma_1 = \frac{p_{\text{u}}}{\sigma^2}\sum_{l=0}^{L-1}\sigma_{h1l}^2$. The PDP of the channel is $\sigma_{hkl}^2 = 1/L$, $l = 0, 1, \ldots, L-1$, and $k = 1, 2, \ldots, K$. We also plot the CRLB (given by \eqref{eq:crlb}). We have $N = 100$ and $L = 2$. We plot both the theoretical MSE (see \eqref{eq:msewk}) and the simulated MSE. With $M = 40$ and $K = 5$ (i.e. massive MIMO regime), it is observed that the proposed estimator is near optimal at high SNR (i.e. the value of SNR for which MSE $\ll \omega_k^2$). However with $M = 2$ and $K = 2$ (i.e. non-massive MIMO regime), we note that the proposed estimator does not perform as good as the CRLB. The above observations therefore motivate the use of the proposed estimator for massive MIMO systems. This is even more so because the well-known near-optimal estimators used in conventional small MIMO systems have prohibitive complexity and cannot be used in massive MIMO systems. On the contrary the proposed estimator has \textit{low-complexity} and is therefore well-suited for massive MIMO regime.

\par Also, for $M = 40$ and $K = 5$, since $G_k \approx 1$ (see Remark $4$), we have $\gamma_1^{0} \approx 0.0055$ (i.e. $-22.5$ dB). As can be seen in Fig.~\ref{fig:crlbmse}, for all $\gamma_k \gg \gamma_k^{0} = -22.5$ dB, the simulated MSE (curve marked with triangles) and the theoretical MSE (curve marked with squares) are exactly the same, i.e., the MSE approximation in \eqref{eq:msewk} is tight. \hfill \qed

\par Using the Corollary to Theorem $1$, we present the following propositions.

\indent \textbf{{Proposition 1:\label{snrk}}} If $|\omega_k KL| \ll \pi$ and $\gamma_k \gg \gamma_k^{0}$, then for a desired MSE, $\epsilon >0$, and fixed $M$, $K$, $L$ and $N$, the required received SNR, $\gamma_k$ is given by\footnote[7]{For ease of notation, subsequently we would be treating the R.H.S. of \eqref{eq:msewk} as the exact value for MSE since the approximation in \eqref{eq:msewk} is tight when $|\omega_k KL| \ll \pi$ and $\gamma_k \gg \gamma_k^{0}$.} 

\vspace{-0.5 cm}

\begin{IEEEeqnarray}{rCl}
\label{eq:snrkexact}
\gamma_k(\epsilon) & \Define & \dfrac{G_k/(B-1)}{2\epsilon d}  \bigg[1 + \sqrt{1 + \dfrac{2(B-1)^2\epsilon d}{KG_k^2}}\bigg]
\IEEEeqnarraynumspace
\end{IEEEeqnarray}

\noindent where $d \Define M(N-KL)(KL)^2G_k^2$.

\begin{IEEEproof}
Since $|\omega_k KL| \ll \pi$ and $\gamma_k \gg \gamma_k^{0}$, then from \eqref{eq:msewk}, for a desired MSE $\epsilon>0$, we get

\vspace{-0.5 cm}

\begin{IEEEeqnarray}{lCl}
\label{eq:quaeqn}
\nonumber \epsilon  =  \dfrac{\dfrac{1}{\gamma_k}\bigg(\dfrac{G_k}{B-1} + \dfrac{1}{2K\gamma_k}\bigg)}{M(N-KL)(KL)^2G_k^2}\\
\text{or, } \epsilon d \gamma_k^2 - \frac{G_k}{B-1} \gamma_k - \dfrac{1}{2K}& = & 0,
\IEEEeqnarraynumspace
\end{IEEEeqnarray}

\noindent where $d = M(N-KL)(KL)^2G_k^2$ and $N = BKL$. Solving \eqref{eq:quaeqn} for the required $\gamma_k$, we obtain \eqref{eq:snrkexact}. \hfill \IEEEQEDhere
\end{IEEEproof}

\textit{Remark 7:\label{snrexact}} From \eqref{eq:snrkexact} it follows that for a fixed $M$, $N$, $K$ and $L$, the required $\gamma_k(\epsilon)$ increases with decreasing desired MSE, $\epsilon$. Further it can be shown that to achieve a desired MSE $\epsilon \ll \omega_k^2$, the required $\gamma_k(\epsilon)$ is much greater than $\gamma_k^{0}$, i.e., $\gamma_k(\epsilon) \gg \gamma_k^{0}$.  \hfill \qed

\indent \textbf{{Proposition 2:\label{varM}}} (Impact of the Number of BS Antennas) Consider $|\omega_k KL|\ll \pi$ and $\gamma_k \gg \gamma_k^{0}$. For a fixed $K$, $L$, $N$ and $M$ sufficiently large, i.e.,

\vspace{-0.5 cm}

\begin{IEEEeqnarray}{rCl}
\label{eq:condM}
M & \gg & \frac{K}{2\epsilon (N-KL)^3},
\IEEEeqnarraynumspace
\end{IEEEeqnarray}

\noindent the required SNR to achieve a fixed desired MSE $\epsilon \ll \omega_k^2$ is inversely proportional to $\sqrt{M}$, i.e., $\gamma_k(\epsilon) \propto 1/\sqrt{M}$.\footnote[8]{Even with increasing $M$ the required $\gamma_k(\epsilon) \gg \gamma_k^{0}$. This follows from Remark $7$ and the fact that $\gamma_k^{0} \propto 1/\sqrt{M}$ for fixed $N$, $K$ and $L$ (see the expression for $\gamma_k^{0}$ in Theorem $1$).}

\begin{IEEEproof}
From \eqref{eq:condM}, we have

\vspace{-0.5 cm}

\begin{IEEEeqnarray}{lCl}
\label{eq:Mapprox}
2\epsilon(N-KL)^3M & \gg & K \implies \dfrac{2(B-1)^2\epsilon d}{KG_k^2}  \gg  1
\IEEEeqnarraynumspace
\end{IEEEeqnarray}

\vspace{-0.2 cm}

\noindent where $d = M(N-KL)(KL)^2G_k^2$ and $N = BKL$. From \eqref{eq:Mapprox} it follows that

\vspace{-0.4 cm}

\begin{IEEEeqnarray}{rCl}
\label{eq:cond2}
\left[1 + \sqrt{1 + \dfrac{2(B-1)^2\epsilon d}{KG_k^2}}\,\, \right] \approx \sqrt{\dfrac{2(B-1)^2\epsilon d}{KG_k^2}}.
\IEEEeqnarraynumspace
\end{IEEEeqnarray}

 \indent Using the approximation from \eqref{eq:cond2} in \eqref{eq:snrkexact}, we get

\vspace{-0.4 cm}

\begin{IEEEeqnarray}{rCl}
\label{eq:mseM}
\gamma_k(\epsilon) \approx \dfrac{G_k}{2\epsilon d}\sqrt{\dfrac{2\epsilon d}{K G_k^2}} \myd \dfrac{1/\sqrt{M}}{\sqrt{2\epsilon K^3L^2(N-KL)G_k^2}},
\IEEEeqnarraynumspace
\end{IEEEeqnarray}

\vspace{-0.15 cm}

\noindent where $(a)$ follows from $d = M(N-KL)(KL)^2G_k^2$. It is clear from \eqref{eq:mseM} that with fixed $K$, $L$ and $N$, for a desired MSE, $\epsilon$, $\gamma_k(\epsilon)$ decreases as $1/\sqrt{M}$ with increasing $M$. \hfill \IEEEQEDhere
\end{IEEEproof}

\textit{Remark 8:\label{snrM}} From proposition $2$ it follows that for a fixed $N$, $K$, $L$ and a fixed desired MSE $\epsilon$, the required SNR $\gamma_k(\epsilon)$ decreases by approximately $1.5$ dB with every doubling in the number of BS antennas, $M$. The same behaviour is exhibited by the numerically simulated required SNR (see Fig.~\ref{fig:varM}). We explain this observation in the following. For a fixed $N$, $K$ and $L$ and a fixed desired MSE, $\epsilon$, the variance of $\nu_k^Q$ must also be fixed (see \eqref{eq:varerr}). The variance of $\nu_k^Q$ is given by the R.H.S. of \eqref{eq:vkvarQ} which consists of two terms, the first term is proportional to $1/M\gamma_k$ and the second term is proportional to $1/M\gamma_k^2$. Therefore, for a fixed variance of $\nu_k^Q$, $\gamma_k$ cannot be reduced faster than $1/\sqrt{M}$ with increasing $M$, since otherwise the second term will increase with increasing $M$. \hfill \qed

\textit{Remark 9:\label{snrK}} When $KL \ll N$ and $M$ is sufficiently large (see \eqref{eq:condM}), the required SNR $\gamma_k(\epsilon)$ to achieve a desired MSE $\epsilon$ is approximately given by (replacing $(N-KL)$ by $N$ in \eqref{eq:mseM})

\vspace{-0.45 cm}

\begin{IEEEeqnarray}{rCl}
\label{eq:snrK}
\gamma_k(\epsilon) & \approx & \dfrac{1}{K^{3/2}}\dfrac{1}{\sqrt{2\epsilon MNL^2G_k^2}}.
\end{IEEEeqnarray}

\indent From \eqref{eq:snrK} it is clear that with fixed $M$, $L$, $N$, fixed desired MSE $\epsilon$ and with increasing $K$, the required SNR $\gamma_k(\epsilon)$ decreases with increasing $K$ as long as $KL \ll N$.

\par In the following we explain why the MSE decreases with increasing $K$ (fixed $M$, $L$ and $N$). From \eqref{eq:cfoest} we note that the estimate $\hat{\omega}_k$ is $1/KL$ times $\arg(\rho_k)$. The division by $K$ helps in reducing the impact of the noise term in $\rho_k$ (i.e. $\nu_k$, see \eqref{eq:cfonoise}) on the MSE of the proposed estimate. Note that the division of $\arg(\rho_k)$ by $KL$ is only  because the proposed \textit{pilot-blocks} are spaced $KL$ channel uses apart.\footnote[9]{Note that in \eqref{eq:eqnintmed1}, the correlation $r_m^{\ast}[\tau(b,k,l)]r_m[\tau(b+1,k,l)]$ between adjacent \textit{pilot-blocks} lead to the term $KLp_{\text{u}}|h_{km}[l]|^2e^{j\omega_k KL}$.} Further the variance of the noise term in $\rho_k$ (i.e. $\nu_k$) is also a function of $K$ (see \eqref{eq:vkvarI} and \eqref{eq:vkvarQ}). In both \eqref{eq:vkvarI} and \eqref{eq:vkvarQ}, both the terms on the R.H.S. have $MKL(B-1) = M(N-KL)$ in the denominator. For $KL \ll N$, $(N - KL) \approx N$ and therefore the first term on the R.H.S. of both \eqref{eq:vkvarI} and \eqref{eq:vkvarQ} does not vary significantly with increasing $K$. However the second term on the R.H.S. has $1/2K\gamma_k^2$ in the numerator which decreases with increasing $K$.\footnote[10]{Note that the second term in the R.H.S. of \eqref{eq:vkvarI} and \eqref{eq:vkvarQ} is proportional to $1/K$ due to the fact that the pilot used is an impulse which is transmitted once every $KL$ channel uses and therefore its instantaneous power is $KL$ times higher than the average transmitted power, $p_{\text{u}}$ (see \eqref{eq:kpilotmat}).} Therefore the variance of the noise term in $\rho_k$ also decreases with increasing $K$ as long as $KL \ll N$. Finally the division of $\arg(\rho_k)$ by $K$ and the reduction in the variance of $\nu_k$ with increasing $K$, results in reduction of the MSE of $\hat{\omega}_k$ with increasing $K$. This automatically implies that for a desired MSE $\epsilon$ the required SNR $\gamma_k(\epsilon)$ decreases with increasing $K$. The same behaviour is also exhibited by the numerically simulated required SNR (see Fig.~\ref{fig:varK}).

\par In Remark $3$ we have seen that the total complexity of the proposed CFO estimator does not increase with increasing $K$ ($N \gg K$). Therefore, with increasing $K$ ($N \gg KL$), the required SNR decreases with no increase in complexity, which is interesting. \hfill \qed

\section{Numerical Results and Discussions}

In all simulation studies presented in this section, we assume an operating carrier frequency $f_c = 2$ GHz and a maximum CFO of $0.1$ PPM of $f_c$ (see footnote $6$ in Remark $1$). Therefore $|\omega_k| \leq \frac{\pi}{2500}$. The communication bandwidth is $B_{\text{w}} = 1$ MHz, the maximum delay spread is $T_{\text{d}} = 5 \mu$s and the coherence interval is $T_c = 1$ ms. Therefore from the proposed communication strategy in Section II (see Fig.~\ref{fig:uplink}), we have $N = 500$ (since $N = N_c/2$ and $N_c = T_c B_{\text{w}} = 1000$) and, $L = T_{\text{d}}B_{\text{w}} = 5$. The PDP is the same for each user and is given by $\sigma_{hkl}^2 = 1/L$, $l = 0, 1, \ldots, L-1$, $k = 1, 2, \ldots, K$. Subsequently for the first user (i.e. $k = 1$) we present the variation in the required SNR (to achieve a fixed desired MSE) as a function of increasing $M$ and $K$.

\par
\begin{figure}[t]
\hspace{-0.5 in}
\includegraphics[width= 4 in, height= 2.5 in]{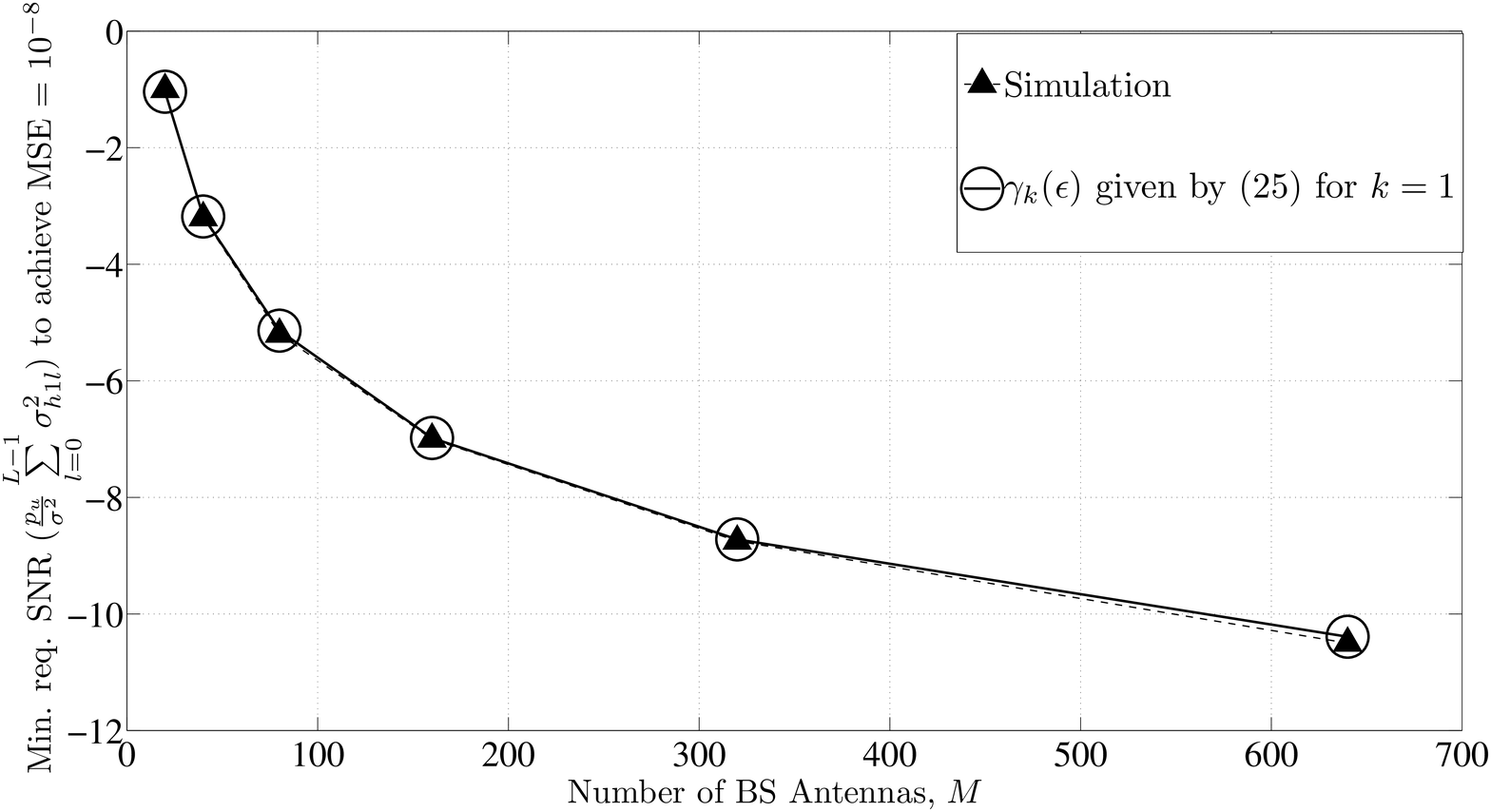}
\caption {Plot of the required SNR to achieve a fixed desired MSE, $\epsilon = \E[(\hat{\omega}_1 - \omega_1)^2] = 10^{-8}$ with increasing number of BS antennas, $M$, for the following fixed parameters: $N = 500$, $K = 10$, $L = 5$, $\omega_1 = \frac{\pi}{2500}$.}
\label{fig:varM}
\end{figure}

 In Fig.~\ref{fig:varM} for a fixed $N$, $L$, $K = 10$ and fixed desired MSE $\epsilon = 10^{-8} \ll \omega_1^2$ ($\omega_1 = \pi/2500$), we plot the required SNR $\frac{p_{\text{u}}}{\sigma^2}\sum_{l=0}^{L-1}\sigma_{hkl}^2$ as a function of increasing $M$. We plot both the numerically simulated required SNR as well as the analytical expression for the required SNR from \eqref{eq:snrkexact}. From Fig.~\ref{fig:varM} it can be seen that the analytical expression in \eqref{eq:snrkexact} is a tight approximation to the exact simulated required SNR. It is also observed that for sufficiently large $M$, the required SNR decreases by roughly $1.5$ dB with every doubling in $M$ as is suggested by Proposition $2$ (see also Remark $8$).

\par In Fig.~\ref{fig:varK} for a fixed $M = 160$, fixed $N$, $L$ and a fixed desired MSE $\epsilon = 10^{-8} \ll \omega_1^2$ ($\omega_1 = \pi/2500$), we plot the numerically simulated required SNR $\frac{p_{\text{u}}}{\sigma^2}\sum_{l=0}^{L-1}\sigma_{hkl}^2$ as a function of increasing $K$. It is also observed that the analytical approximation to the required SNR (see \eqref{eq:snrkexact}) is tight. As discussed in Remark $9$, the required SNR decreases with increasing number of UTs, $K$.

\appendix[Proof of Theorem 1]

With $\nu_k^I \Define \Re(\nu_k)$ and $\nu_k^Q \Define \Im(\nu_k)$, from \eqref{eq:rhokredef} we get

\vspace{-0.5 cm}

\begin{IEEEeqnarray}{rCl}
\label{eq:argpk}
\arg (\rho_k) = \tan^{-1} \bigg [\dfrac{G_k\sin(\omega_k KL) + \nu_k^Q}{G_k\cos(\omega_k KL) + \nu_k^I}\bigg ].
\IEEEeqnarraynumspace
\end{IEEEeqnarray}

\vspace{-0.2 cm}

\par Note that $|\omega_k KL| \ll \pi \implies \cos(\omega_k KL) \approx 1$. Using this approximation in \eqref{eq:vkvarI}, we get

\vspace{-0.5 cm}

\begin{IEEEeqnarray}{rCl}
\label{eq:spcond1}
\dfrac{\E[(\nu_k^I)^2]}{G_k^2} & \approx & \frac{\dfrac{(2B-3)G_k}{(B-1)\gamma_k} + \frac{1}{2K\gamma_k^2}}{M(B-1)KLG_k^2} \myg 1,
\IEEEeqnarraynumspace
\end{IEEEeqnarray}

\vspace{-0.2 cm}

\noindent where $(a)$ follows from $\gamma_k \gg \gamma_k^{0}$. Therefore we conclude that $\nu_k^I/G_k$ is small (i.e., $\frac{\nu_k^I}{G_k} \ll 1$) with high probability. Using this approximation, we can write $G_k \cos(\omega_k KL) + \nu_k^I \approx G_k\cos(\omega_k KL)$. Since $\E[(\nu_k^Q)^2] < \E[(\nu_k^I)^2]$ (compare \eqref{eq:vkvarI} and \eqref{eq:vkvarQ} with the approximation $\cos(\omega_k KL) \approx 1$), we can also say that $\frac{\nu_k^Q}{G_k} \ll 1$ with high probability. Using above approximations, in \eqref{eq:argpk}, we get

\vspace{-0.5 cm}

\begin{IEEEeqnarray}{rCl}
\label{eq:argpk1}
\nonumber \arg(\rho_k) & \approx & \tan^{-1}\bigg[ \tan(\omega_k KL) + \dfrac{\nu_k^Q}{G_k\cos(\omega_k KL)}\bigg]\\
\nonumber & \mya & \tan^{-1}\bigg[ \tan(\omega_k KL) + \dfrac{\nu_k^Q}{G_k}\bigg]\\
\nonumber & \myb & \tan^{-1}\left[\dfrac{\tan(\omega_k KL) + \tan(\dfrac{\nu_k^Q}{G_k})}{1 - \tan(\omega_k KL)\tan(\dfrac{\nu_k^Q}{G_k})}\right]\\
& \myc & \tan^{-1}\bigg[\tan \bigg(\omega_k KL + \dfrac{\nu_k^Q}{G_k}\bigg)\bigg] = \omega_k KL + \frac{\nu_k^Q}{G_k},
\IEEEeqnarraynumspace
\end{IEEEeqnarray}

\noindent where $(a)$ follows from substituting $\cos(\omega_k KL) \approx 1$. Step $(b)$ follows from the fact that since $\frac{\nu_k^Q}{G_k} \ll 1$ with high probability, $\frac{\nu_k^Q}{G_k} \approx \tan(\frac{\nu_k^Q}{G_k})$ and $|\tan(\omega_k KL)\tan(\frac{\nu_k^Q}{G_k})| \ll 1$ with high probability. Step $(c)$ follows from the standard result $\tan(A + B) = (\tan A + \tan B)/(1 - \tan A \tan B)$. Substituting \eqref{eq:argpk1} in \eqref{eq:cfoest}, we get the desired approximation for the CFO estimate in \eqref{eq:approxest1}.


\begin{figure}[t]
\hspace{-0.5 in}
\includegraphics[width= 4 in, height= 2.5 in]{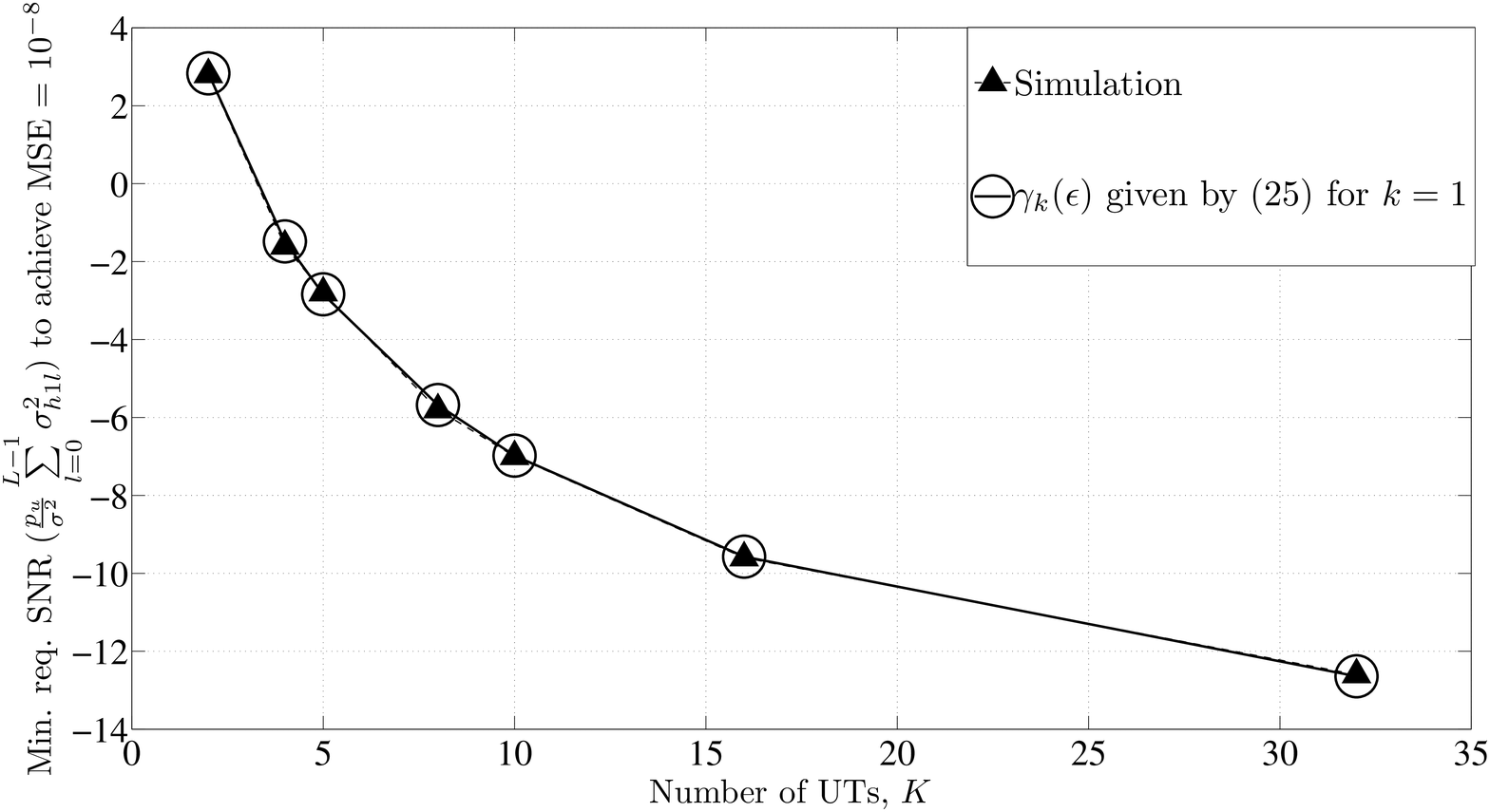}
\caption {Plot of the required SNR to achieve a fixed desired MSE, $\epsilon = \E[(\hat{\omega}_1 - \omega_1)^2] = 10^{-8}$ with increasing number of UTs, $K$, for the following fixed parameters: $N = 500$, $M = 160$, $L = 5$, $\omega_1 = \frac{\pi}{2500}$.}
\label{fig:varK}
\end{figure}



%

%


\ifCLASSOPTIONcaptionsoff
  \newpage
\fi



%

\bibliographystyle{IEEEtran}
\bibliography{IEEEabrvn,mybibn}

\begin{thebibliography}{10}
\providecommand{\url}[1]{#1}
\csname url@samestyle\endcsname
\providecommand{\newblock}{\relax}
\providecommand{\bibinfo}[2]{#2}
\providecommand{\BIBentrySTDinterwordspacing}{\spaceskip=0pt\relax}
\providecommand{\BIBentryALTinterwordstretchfactor}{4}
\providecommand{\BIBentryALTinterwordspacing}{\spaceskip=\fontdimen2\font plus
\BIBentryALTinterwordstretchfactor\fontdimen3\font minus
  \fontdimen4\font\relax}
\providecommand{\BIBforeignlanguage}[2]{{%
\expandafter\ifx\csname l@#1\endcsname\relax
\typeout{** WARNING: IEEEtran.bst: No hyphenation pattern has been}%
\typeout{** loaded for the language `#1'. Using the pattern for}%
\typeout{** the default language instead.}%
\else
\language=\csname l@#1\endcsname
\fi
#2}}
\providecommand{\BIBdecl}{\relax}
\BIBdecl

\bibitem{Andrews}
J.~Andrews, S.~Buzzi, W.~Choi, S.~Hanly, A.~Lozano, A.~Soong, and J.~Zhang,
  ``What {W}ill 5{G} {B}e?'' \emph{{IEEE} J. Sel. Areas Commun.}, vol.~32,
  no.~6, pp. 1065--1082, June 2014.

\bibitem{Marzetta1}
T.~Marzetta, ``Noncooperative {C}ellular {W}ireless with {U}nlimited {N}umbers
  of {B}ase {S}tation {A}ntennas,'' \emph{{IEEE} Trans. Wireless Commun.},
  vol.~9, no.~11, pp. 3590--3600, November 2010.

\bibitem{Ngo1}
H.~Q. Ngo, E.~Larsson, and T.~Marzetta, ``Energy and {Spectral} {Efficiency} of
  {Very} {Large} {Multiuser} {MIMO} {Systems},'' \emph{{IEEE} Trans. Commun.},
  vol.~61, no.~4, pp. 1436--1449, April 2013.

\bibitem{Stoica}
P.~Stoica and O.~Besson, ``Training sequence design for frequency offset and
  frequency-selective channel estimation,'' \emph{{IEEE} Trans. Commun.},
  vol.~51, no.~11, pp. 1910--1917, Nov 2003.

\bibitem{Ma}
J.~Chen, Y.-C. Wu, S.~Ma, and T.-S. Ng, ``Joint {CFO} and {C}hannel
  {E}stimation for {M}ultiuser {MIMO-OFDM} {S}ystems with {O}ptimal {T}raining
  {S}equences,'' \emph{{IEEE} Trans. Signal Process.}, vol.~56, no.~8, pp.
  4008--4019, Aug 2008.

\bibitem{Simon}
E.~P. Simon, L.~Ros, H.~Hijazi, and M.~Ghogho, ``Joint {C}arrier {F}requency
  {O}ffset and {C}hannel {E}stimation for {OFDM} {S}ystems via the {EM}
  {A}lgorithm in the {P}resence of {V}ery {H}igh {M}obility,'' \emph{{IEEE}
  Trans. Signal Process.}, vol.~60, no.~2, pp. 754--765, Feb 2012.

\bibitem{Ghogho}
M.~Ghogho and A.~Swami, ``Training {D}esign for {M}ultipath {C}hannel and
  {F}requency-{O}ffset {E}stimation in {MIMO} {S}ystems,'' \emph{{IEEE} Trans.
  Signal Process.}, vol.~54, no.~10, pp. 3957--3965, Oct 2006.

\bibitem{Poor}
Y.~Yu, A.~Petropulu, H.~Poor, and V.~Koivunen, ``Blind estimation of multiple
  carrier frequency offsets,'' in \emph{Personal, Indoor and Mobile Radio
  Communications, 2007. PIMRC 2007. IEEE 18th International Symposium on}, Sept
  2007, pp. 1--5.

\bibitem{Larsson2}
H.~Cheng and E.~Larsson, ``Some {F}undamental {L}imits on {F}requency
  {S}ynchronization in massive {MIMO},'' in \emph{Signals, Systems and
  Computers, 2013 Asilomar Conference on}, Nov 2013, pp. 1213--1217.

\bibitem{Weiss}
M.~Weiss, ``Telecom {R}equirements for {T}ime and {F}requency
  {S}ynchronization,,'' National Institute of Standards and Technology (NIST),
  USA, [Online]: www.gps.gov/cgsic/meetings/2012/weiss1.pdf.

\end{thebibliography}

\end{document}